\documentstyle[sprocl]{article}            % For archive submission

\def\Journal#1#2#3#4{{#1} {\bf #2}, #3 (#4)}

\def\NPB{{\em Nucl. Phys.} B}
\def\NPBPS{{\em Nucl. Phys.} B(Proc. Suppl.)}
\def\PLB{{\em Phys. Lett.} B}
\def\PLA{{\em Phys. Lett.} A}

\def\RMP{\em Rev. Mod. Phys.}
\def\PRD{{\em Phys. Rev.} D}
\def\CMP{\em Comm. Math. Phys.}
\def\EPJC{{\em Eur. Phys. J.} C}

\newcommand{\beq}{\begin{equation}}
\newcommand{\eeq}{\end{equation}}
\newcommand{\fie}{\varphi}
\newcommand{\half}{{\scriptstyle{{1\over 2}}}}
\newcommand{\Tr}{\mbox{\,Tr\,}}

\begin{document}
${}$\vskip-18mm\hfill\vbox{\hbox{FTUAM-00-13}              % preprint ON
                           \hbox{INLO-PUB-7/00}}\vskip9mm % preprint ON
\title{OVERLAPPING INSTANTONS~\footnote{Presented by 
PvB at ``Continuous advances                               % preprint ON
in QCD, IV'', Minneapolis, 12-14 May, 2000.}               % preprint ON
%Pierre van Baal}                                             % preprint OFF
}
\author{Margarita Garc\'{\i}a P\'erez,}
\address{Departamento de F\'{\i}sica Te\'orica, C-XI, Universidad Aut\'onoma 
de Madrid,\\ Madrid 28049, Spain}
\author{Tam\'as G. Kov\'acs and Pierre van Baal}
\address{Instituut-Lorentz for Theoretical Physics, University of Leiden,\\
P.O. Box 9506, Nl-2300 RA Leiden, The Netherlands} 
\maketitle
\abstracts{Overlapping instantons have an action density profile that 
significantly deviates from the simple addition of the density profiles
of single instantons. This turns out to have important consequences for
identifying the proper instanton content of a given configuration. Most
dramatic is the case where the instantons are parallel in group space, 
leading to the effect of hiding large instantons. Sufficiently large 
instantons can have important contributions to a confining interaction.
}
\centerline{\em DEDICATED TO THE MEMORY OF MICHAEL MARINOV}
\section{Introduction}

In recent years the instanton liquid model has been very successful in 
describing the low energy properties of light hadrons~\cite{Shuryak}. 
Whether or not instantons can account for confinement has recently been 
much discussed~\cite{Conf} again. Relevant for this discussion is if 
there are sufficiently many large (say bigger than $1/\Lambda_{\rm QCD}$)
instantons. Once such large instantons appear, it has been argued that 
all types of topological excitations should show up, more or less on equal 
footing~\cite{Lat97}. Indeed, recent studies on calorons, which can be 
viewed as overlapping instantons when their size becomes bigger than the
inverse temperature, have shown how instantons and monopoles are intimately 
connected~\cite{Thomas}. One difficulty is that one should not expect to be 
able to account for large instantons using semiclassically inspired techniques. 

Starting from an effective action that incorporates the $\theta$ angle 
in a manner consistent with all known Ward identities, has led naturally 
to a Coulomb gas representation in terms of fractionally charged 
objects~\cite{Zhit}, that bear striking resemblance with the ``instanton 
quarks'' coined for describing the semiclassical parameters~\cite{I-qs}, and 
with the more tangible constituent monopoles at finite 
temperature~\cite{Thomas}. The description is very inspiring and has the 
advantage that it does not rely on semiclassical considerations. 

So what can be learned from lattice simulations~\cite{Lat,I1,I2} about the
presence of large instantons? Typically these simulations are done on too 
small volumes and have too poor statistics to contain any precise information 
on the tail of the distribution. One should thus not be too hasty in declaring 
it a fact that large instantons are (exponentially) suppressed~\cite{Fact}. 
Apart from the finite volume cutoff, we find there is an intrinsic difficulty 
in identifying large instantons, due to overlapping effects. In these 
situations the usual assumption that the liquid is dilute enough, i.e. the 
individual pseudoparticles are far enough apart that they do not distort one 
another considerably, is no longer valid. The notion of ``instanton size'' now 
becomes ambiguous, even when ignoring the influence of quantum fluctuations. 
It should be said that no definition of ``instanton size'' in terms of 
physical observables is known.

Even considering exact charge 2 solutions, within a semiclassical context, 
we show~\cite{Us} that the correspondence between the instanton parameters 
and single instanton sizes is not always well-defined. Despite its limitations,
the results are so simple, and its consequences so surprising that it is clear 
this issue needs to be addressed further in order to understand which field 
configurations are important for the long distance features of QCD.

\section{Parallel gauge orientation}

To demonstrate the point of non-linear effects for overlapping instantons most 
strongly, we look at the simple case of an exact charge 2 instanton solution 
with instanton constituents that are parallel in group space. These can be 
described in a straightforward way using the 't~Hooft ansatz~\cite{Hooft},
$A_\mu=\half\bar\eta_{\mu\nu}\partial_\nu\log(\phi(x))$, with $\phi^{-1}
\partial_\mu^2\phi=0$. Here $\bar\eta_{\mu\nu}=i\tau_a\bar\eta^a_{\mu\nu}=
\bar\sigma_{[\mu}\sigma_{\nu]}$, where $\bar\sigma_\mu=\sigma_\mu^\dagger$
are unit quaternions in the $2\times2$ matrix representation $\sigma_0=I_2$
and $\sigma_a=i\tau_a$ ($\tau_a$ the Pauli matrices). Charge 2 instanton 
solutions are described by
\beq
\phi(x)=1+\frac{\rho_1^2}{(x-a)^2}+\frac{\rho_2^2}{(x-b)^2},
\eeq
where $\rho_1$ and $\rho_2$ are the sizes of the two instantons, one at 
$x=a$ and the other located at $x=b$. These instantons will start to 
overlap when $(a-b)^2\sim \rho_i^2$. For any $a\neq b$ the two poles of 
$\phi$ reflect the fact that the charge is 2. {\em But} when $a=b$ we seem 
to be left with one single pole, and thus with a charge 1 instanton of size 
$\sqrt{\rho_1^2+\rho_2^2}$. {\em How can this be}?
The answer is that the charge 2 instanton, when its two constituents
overlap, is far from looking like the superposition of two charge 1 instantons
(summing the action densities of two single instantons with the same 
parameters). Actually, it looks like a narrow instanton on top of a broad 
instanton. When $\rho_1=\rho_2=\rho$, the narrow instanton has a size 
given by half the distance $|a-b|$ and the broad instanton has a size 
$\sqrt{2}\rho$. 
It is the broad instanton that is left over in the limit $a\rightarrow b$,
whereas the narrow instanton becomes singular. It forms the boundary of
the moduli space and as such is not new~\cite{Don}. But what 
this means in terms of how the configurations {\em look} like, when 
approaching this boundary, was not investigated in the past.

\begin{figure}[htb]
\vspace{10.6cm}
\includegraphics{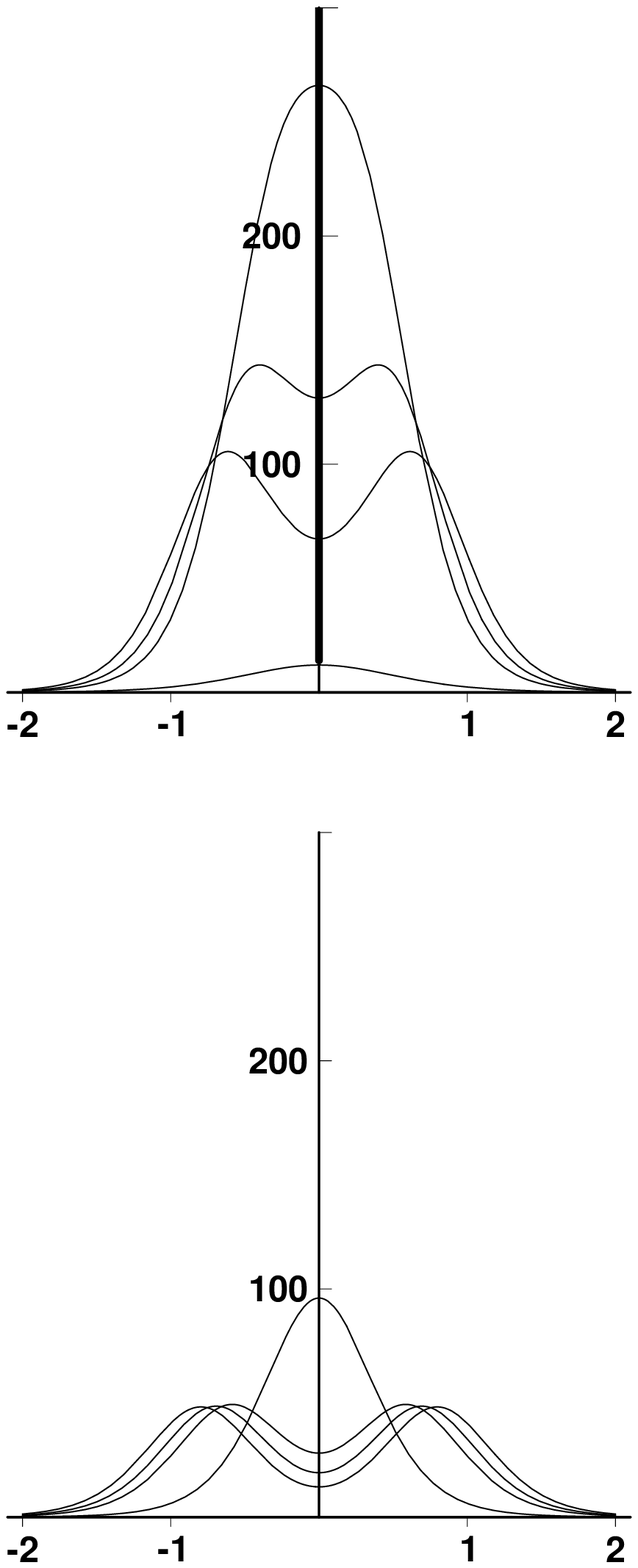}
\includegraphics{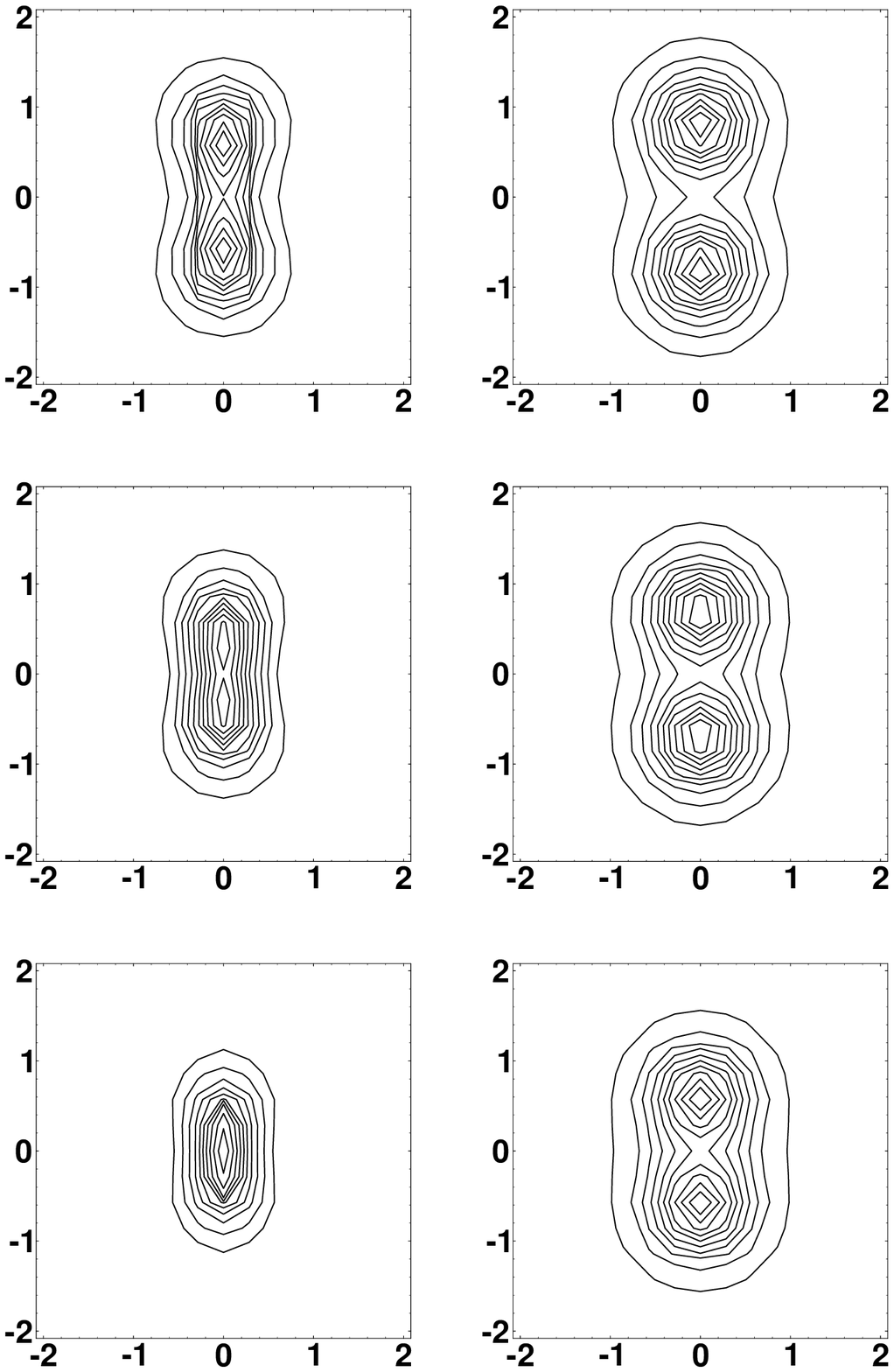}
\caption{Left plot shows the profile along the line connecting to equal size 
instantons (scaled to 1) at separations 0.0, 1.2, 1.4 and 1.6. Top is for the 
exact charge 2 solution (at zero separation the fat line indicates the 
singular instanton on a $\protect\sqrt{2}$ size instanton), bottom for the 
superposition of two charge 1 instantons. On the right is shown the contour 
plots for the same parameters, from top to bottom at decreasing separation 
(not showing zero separation). Left are the contours for the exact charge 2 
solution and right those based on superposition.}
\label{fig:profile}
\end{figure}

We demonstrate this simple observation in two figures that show the action 
density. We plot in fig.~\ref{fig:profile} both the action density along the 
line connecting the two centers, labelled as $a=y+z$ and $b=y-z$, and as 
contour plots including a perpendicular direction (there is an $O(3)$ `axial' 
symmetry), for a number of $|z|$ values (choosing $\rho_1=\rho_2=1$). Both 
are compared to what one would get from simply superposing two charge 1 
instantons. These plots are generated using the simple formula for the 
action density (subtracting the delta function singularities at $x=a$
and $x=b$)
\beq
s(x)=-\half\Tr F_{\mu\nu}^2(x)=-\half\partial_\mu^2\partial_\nu^2\log(\phi(x)).
\label{eq:action}
\eeq
We see that for $|z|$ of the order of $\rho$ the exact solution rises 
considerably over the superposition result. As the total action is 
in both cases of course equal (to $16\pi^2$), this is compensated by
a considerable narrowing of the configuration in the transverse direction.
At the moment where the two peaks in the exact charge 2 density profile
merge (at $|z|=\sqrt{0.4}\rho$ for the case $\rho_1=\rho_2=\rho$), very
soon the narrow instanton starts to dominate and becomes $O(4)$ symmetric,
on the background of the broad $O(4)$ symmetric instanton that is left
for $z=0$.

It has the dramatic consequence of hiding large instantons, those with
sizes comparable to, or larger than the average instanton distance. It 
may therefore, even in large volumes, explain the observed exponential
fall off for the instanton size distribution, extracted from the lattice
data~\cite{Lat,I1,I2}. This is not only because in the lattice studies
one does assume one can approximately describe the (smoothed) configurations
in terms of superpositions of single (anti-)instantons, it is also because
there is an ambiguity in the parametrisation. One either has two large
instantons or one small instanton on top of an even larger one. 

We generated the charge density (here equal to the action density) of a set 
of instanton pairs with the instanton scale parameters $(\rho_{1,2})$ 
distributed independently and qualitatively similar to that found on the 
lattice. We artificially enhanced the tail of the distribution in order to 
test whether such a tail can remain undetected by the lattice instanton 
finders. The separation $2|z|$ was Gaussian distributed with mean 7.0, and 
variance of 1.0. The resulting charge densities --- each containing one pair 
resolved on a $16^4$ grid --- were then analysed using two different instanton 
finder algorithms~\cite{I1,I2}. The details of these algorithms are not 
relevant in the present context. However their most important common feature 
is that they are both based on the dilute gas assumption. They identify the 
highest peaks in the charge density and estimate the instanton sizes from 
the fall-off of the density in the vicinity of the maximum. Only when the 
individual pseudoparticles are far enough apart that they do not distort 
one another considerably, does the ``instanton size'' have an unambiguous 
meaning. Treating the charge 2 case exactly can be thought of as the next 
order approximation when one takes into account the distorting effect of 
like charge nearest neighbour pairs.

\begin{figure}[htb]
\vspace{7.8cm}
\includegraphics{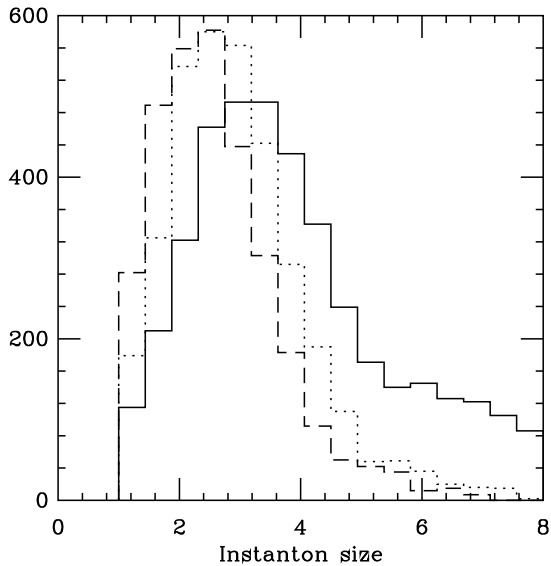}
\caption{The instanton size distribution with the relative gauge 
orientation parallel. The solid line indicates the distribution of 
the size parameter, the dotted and the dashed lines are the size 
distributions determined by the instanton finder algorithms of 
refs.~\protect\cite{I1} and ref.~\protect\cite{I2} respectively.}
\label{fig:par}
\end{figure}

In fig.~\ref{fig:par} we show the instanton size distributions found 
by the algorithms of ref.~\cite{I1} (dotted line) and ref.~\cite{I2} (dashed 
line) along with the distribution of the size parameters used to construct 
the charge densities (solid line). The two instanton finders both yield 
a significantly suppressed tail.

\section{General case}
In the previous section we restricted our attention to the case where 
the instantons are parallel in group space. We expect overlapping 
effects for non-parallel group orientations to be large as well (like 
for calorons~\cite{Thomas} when the size becomes larger than the inverse 
temperature, giving rise to constituent monopoles). To study the general 
charge 2 instanton solutions one could use the conformal extension of 
the 't~Hooft ansatz~\cite{Hooft} (but not for higher charge). However, 
it is generally very cumbersome to relate its parameters to a physical 
description~\cite{Hart}. We therefore used the Atiyah-Drinfeld-Hitchin-Manin 
(ADHM) construction~\cite{ADHM} for the general charge 2 
solution~\cite{Christ}. 

A convenient explicit form for a charge $Q$ self-dual gauge field 
reads~\cite{CFTG}
\begin{equation}
A_\mu(x)=\frac{1}{2}(1-\lambda F(x)\lambda^\dagger)^{-1}\partial_\nu
\left(\lambda\sigma^\dagger_{[\mu}\sigma^{\vphantom{\dagger}}_{\nu]}F(x)
\lambda^\dagger\right),\quad F^{-1}(x)\equiv\Delta^\dagger(x)\Delta(x),
\end{equation}
where $\lambda=(\lambda_1,\cdots,\lambda_Q)$, forming the first row of 
the $Q\times (1+Q)$ quaternionic matrix $\Delta(x)$. The remainder of 
$\Delta(x)$ forms a $Q\times Q$ matrix $B-I_Q x$, with $B$ symmetric and 
$x=x_\mu \sigma_\mu$ denoting the space-time coordinate. This gauge field
is self-dual if and only if $\Delta(x)$ satisfies the ADHM constraint:
$\Delta^\dagger(x)\Delta(x)$ is proportional to $\sigma_0$ and invertible 
as a real $Q\times Q$ matrix. A redundancy of parameters can be removed 
using the symmetry under which the gauge field remains {\em unchanged}, 
$B\rightarrow TBT^{-1},~\lambda\rightarrow\lambda T^{-1}$, for $T\in O(Q)$.

For charge 2, $\Delta(x)$ can be parametrised as follows
\beq
\Delta(x) = \left( \begin{array}{cc}\lambda_1   &    \lambda_2 \\
                                    y+z-x       &       u      \\
                                    u           &     y-z-x\end{array}\right),
\eeq
where, like $x=x_\mu\sigma_\mu$, $\lambda_{1,2}$, $y$, $z$, and $u$ are 
quaternions. The ADHM constraint now reads
\beq
z^\dagger u-u^\dagger z=\half\left( \lambda_2^\dagger \lambda_1 
- \lambda_1^\dagger \lambda_2 \right)\equiv\Lambda,
     \label{eq:Lambda}
\eeq
introducing $\Lambda$ for notational convenience. This constraint has a one
parameter set of solutions~\cite{Christ} given by
\beq
u=\frac{z \Lambda}{2 |z|^2} + \alpha z,
     \label{eq:u}
\eeq
The redundant real parameter $\alpha$ is removed by the $O(2)$ symmetry 
that leaves the gauge field unchanged, but which {\em does mix} the 
parameters $\lambda_i$, $u$ and $z$. As instantons are identified from 
their action (or charge) density profiles, we first recall the simple 
formula~\cite{Osborn}, 
\beq
s(x)=-\half\Tr F_{\mu\nu}^2(x)=-\half\partial_\mu^2\partial_\nu^2\log 
                        \det(\Delta^\dagger(x) \Delta(x)),
     \label{eq:charge}
\eeq
which agrees with the action density for the special case of the 't~Hooft 
solution, eq.~(\ref{eq:action}), for which $\lambda_1=\rho_1\sigma_0$, 
$\lambda_2=\rho_2 \sigma_0$ and $u=0$ (this indeed solves the ADHM 
constraint, eq.~(\ref{eq:Lambda})). At large separations ($|z|$ large), 
the relative gauge orientation does not play a role, and by insisting
$|\lambda_i|$ describe the sizes of the two well-separated constituents
one puts $\alpha=0$ (this can be imposed by $u_\mu z_\mu=0$). Therefore, 
the general charge 2 solution is described by the following set of 13 
free parameters: $\rho_{1,2}=|\lambda_{1,2}|$, the scale parameters; 
$\lambda_1^\dagger\lambda_2/(|\lambda_1| |\lambda_2|)\in SU(2)$, the 
relative gauge orientation; and $y \pm z$ the location of the constituents. 

However, as has been noted before~\cite{Dorey}, there are generically 
16 points ($\Lambda=0$ and $|u|=|z|$ are degenerate cases) on an $O(2)$ 
orbit satisfying $u_\mu z_\mu=0$. Most are related (like $\lambda_1
\leftrightarrow\lambda_2$ and $z\rightarrow -z$) without affecting the 
interpretation, but {\bf one} non-trivial relation remains 
\begin{equation}
 \left( y, \, z, \, \lambda_1, \, \lambda_2, \,
                      u=\frac{z\Lambda}{2|z|^2} \right) \;\;
 \longrightarrow \;\;
 \left(y, \, u, \,
         \frac{\lambda_1+\lambda_2 }{\sqrt{2}}, \,
         \frac{\lambda_1-\lambda_2 }{\sqrt{2}}, \,
         z \right).
     \label{eq:duality}
\end{equation}
\begin{figure}[htb]
\vspace{8.3cm}
\includegraphics{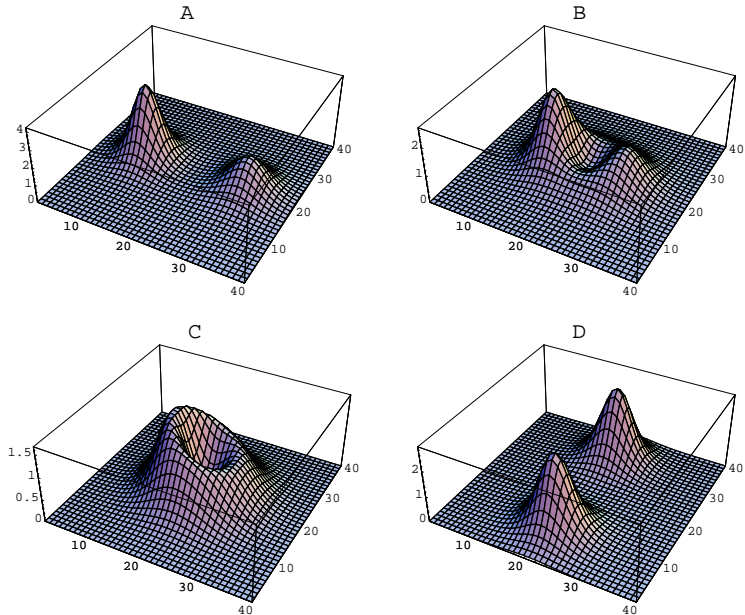}
\caption{A pair of size $\rho_1=6.6$ and $\rho_2=8.3$ instantons
with perpendicular relative gauge orientations. The centres are
separated along the $x_0$ axis, the separation, $2|z|$ is 20.0 (A),
13.3 (B), 10.5 (C), and 5.0 (D). The action density is shown
in the (01) plane.}
\label{fig:scatter}
\end{figure}
This gives rise to a short-to-large distance duality~\footnote{The 
present duality should not be confused with the one described by 
A. Yung~\protect\cite{Yung}, relating a small anti-instanton in the 
background of a large instanton by a conformal transformation to a 
far separated instanton-anti-instanton pair. The gauge field is not
left invariant under this conformal transformation.}, as long as the relative 
gauge orientation is not parallel ($\Lambda \neq 0$). The question now arises 
which of these two descriptions is the ``physical'' one. To answer this it is 
again instructive to {\em look} at the charge density profile of a set of 
solutions with varying separations. In fig.~\ref{fig:scatter} we show such 
a sequence. 

The scale parameters, relative orientation and separation, in terms of the 
l.h.s. of eq.~(\ref{eq:duality}), are described by $\lambda_1 =6.6\sigma_0$, 
$\lambda_2= 8.3\sigma_1$ and $2z=2|z|\sigma_0$. For large $|z|$ (A) the 
constituents are indeed aligned along the 0-axis. As the separation decreases 
the two lumps merge together into an asymmetric ring (B-C). For even smaller 
separation (D) the two lumps separate again but now displaced along the 1-axis.
Clearly in this case the preferred parametrisation is the r.h.s. of 
eq.~(\ref{eq:duality}), describing two instantons with the same scale 
parameter $\rho=7.5$, at a distance of 22. Thus, when $|z|^2\gg|\Lambda|$, 
the original description is ``physical'', i.e. describing two superposed 
instantons separated by a distance $2|z|$. When $|z|^2\ll|\Lambda|$ it is,
however, the dual description which is more ``physical''. 

\begin{figure}[htb]
\vspace{7.8cm}
\includegraphics{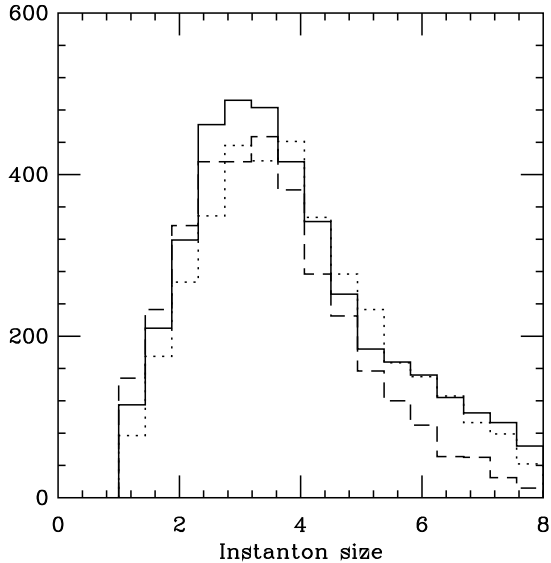}
\caption{The instanton size distribution with the relative orientation 
being distributed according to the Haar measure. The solid line indicates 
the distribution of the ``physical'' (see text) ADHM size parameter, the 
dotted and the dashed lines are the size distributions determined by the 
instanton finder algorithms of ref.~\protect\cite{I1} and 
ref.~\protect\cite{I2} respectively.}
\label{fig:haar}
\end{figure}

In fig.~\ref{fig:par} we plotted the size distributions obtained when all the 
pairs were taken oriented parallel in group space. Following the same procedure
(including the enhanced tail in the generated size distribution), in 
fig.~\ref{fig:haar} we instead consider random colour orientations described 
by the Haar measure. In this case there is no significant suppression of large 
instantons. The ambiguity in the physical parametrisation for non-parallel 
orientation has as a consequence that two instantons can never get closer to 
each other than $2|z|_{min}\equiv\sqrt{2\rho_1\rho_2|\sin\fie|}$, where $\fie$ 
is the invariant angle of the relative group orientation. This seems to have no 
observable effect on the size distribution.
Due to the $\sin^2 \fie$ factor, the Haar measure very strongly favours (close 
to) perpendicular orientation, thus our two distributions almost represent the 
two possible extremes. For the special case of equal size instantons with
perpendicular relative orientation one finds at the minimal distance a 
{\em symmetric} ring, also easily described by the conformal 't~Hooft 
ansatz~\footnote{We thank N. Manton for explaining how also the 
appearance of the {\em asymmetric} ring can be understood from the 
conformal parametrisation~\protect\cite{Hart}.}, by taking 
$\phi=(x-a)^{-2}+(x-b)^{-2}+(x-c)^{-2}$ with $a,b$ and $c$ forming 
an equilateral triangle.

We briefly revisit the case of parallel orientations for which the formalism 
developed in this section is somewhat degenerate. The transition between the 
two parametrisations (eq.~(\ref{eq:duality})) for $\fie\approx 0$ occurs at 
very small separation. In the limit of parallel orientation, one finds
\beq
 \left(y,\,z,\,\rho_1\sigma_0,\,\rho_2\sigma_0,\,0\right)\;\;
 \longrightarrow \;\;\left(y, \, 0, \, 
         \frac{\rho_1+\rho_2 }{\sqrt{2}} \sigma_0, \,
         \frac{\rho_1-\rho_2 }{\sqrt{2}} \sigma_0, \,
         z \right).
     \label{eq:duality1}
\eeq
For $z=0$ the two descriptions are equivalent, and there is no way to 
distinguish between them. One sees that two instantons of scale parameters 
$\rho_{1,2}$ on top of each other is equivalent to a small instanton of size 
$\hat\rho_1=|\rho_2-\rho_1|/\sqrt{2}$ on top of a larger one of size 
$\hat\rho_2=(\rho_2+\rho_1)/\sqrt{2}$. This is consistent with our findings 
in the previous section for $\rho_1=\rho_2$, but it should be noted that for 
$\fie=0$ any choice of instanton sizes $\hat\rho_{1,2}$ are equivalent,
provided $\hat\rho_1^2+\hat\rho_2^2=\rho_1^2+\rho_2^2$. Only {\em looking} 
at the action distribution, see fig.~\ref{fig:profile} and the discussion in
the previous section, tells us what is the ``physical'' choice.

\section{Conclusions}

We have seen how the identification of single instanton parameters becomes 
ambiguous when instantons overlap. We discussed what happens in the two 
cases of parallel and random orientation in group space. We expect that 
the exact way this affects the instanton size distributions measured on 
the lattice will depend on the relative orientation of nearest neighbour 
pairs. 

To summarise, for non-dilute instanton ensembles the next approximation to
a simple superposition is to treat nearest pairs of like charge exactly. 
Staying as close a possible to the dilute picture one is left with two dual 
sets of parameters describing the same charge 2 instanton solution. It 
implies the existence of a minimal distance between the two instantons, 
which is maximal in the case of perpendicular orientation. In the other 
extreme case of (nearly) parallel orientation, two close large instantons
are more naturally described by a small instanton sitting on top of a large 
instanton. Thereby one tends to miss large instantons or to underestimate 
instanton sizes, as was confirmed by a numerical study.  

\section*{Acknowledgements}
Stimulating discussions on dense instanton ensembles with Eric Zhitnitsky, 
on the boundary of moduli space with Francesco Fucito, and on the charge 2 
``ring'' shaped configurations with Nick Manton and Paul Sutcliff are greatly 
appreciated. PvB thanks the organisers for inviting him (for the 3rd time) 
to this 4th workshop on ``Continuous Advances in QCD''. He is also grateful 
to Jac Verbaarschot and the other organisers of the INT-00-1 Program ``QCD at 
Nonzero Baryon Density'' for their invitation, and he thanks the Institute 
for Nuclear Theory at the University of Washington for its hospitality and 
for partial support during the completion of this work. This work was also 
supported in part by a grant from ``Stichting Nationale Computer Faciliteiten 
(NCF)'' for use of the Cray Y-MP C90 at SARA. T. Kov\'acs was supported by 
FOM and M. Garc\'{\i}a P\'erez by CICYT under grant AEN97-1678.

\end{document}